\documentstyle[11pt,newpasp,twoside,psfig]{article} 
\begin{document} 
\title{The 2dF QSO Redshift  Survey.} 
\author{T. Shanks} 
\affil{University of Durham, South Road, Durham, DH1 3LE, UK.} 
\author{B.J. Boyle} 
\affil{AAO, PO Box 296, Epping, NSW 2121, Australia.} 
\author{S.M. Croom} 
\affil{Imperial College, Prince Consort Road, London SW7 1BH, UK} 
\author{N. Loaring, L. Miller} 
\affil{University of Oxford, 1 Keble Road, Oxford, OX1, UK.} \author{R.J. Smith} 
\affil{MSSSO, Private Bag, Weston Creek, ACT 2611, Australia }

\begin{abstract} With  $\approx$6000 QSO redshifts, the 2dF QSO redshift survey is already the
biggest complete QSO survey.  The aim for the survey is  to have 25000 QSO redshifts,  providing an
order of magnitude increase in QSO clustering statistics. We first describe the observational
parameters of the 2dF QSO  survey. We then describe several highlights of the survey so far, 
including  new estimates of the QSO luminosity function and its evolution. We also review the current
status of  QSO clustering analyses from  the 2dF data.  Finally, we  discuss how the complete QSO
survey will be able to constrain the value of $\Omega_o$ by measuring the evolution of QSO
clustering, place limits on the cosmological constant via a direct geometrical test and  determine
the form of the fluctuation power-spectrum out to the $\approx$1000h$^{-1}$Mpc scales only previously
probed by COBE.

\end{abstract}  \section{Introduction}

The observational aim of the 2dF QSO Survey is to use the new AAT 2dF fibre-optic coupler to obtain
redshifts for 25000  B$<$20.85, 0$\la$z$\la$3 QSO's in two $75\times5$deg$^2$ strips of sky  in the
RA ranges  21h50 - 03h15 at $\delta$ = -30$^\circ$ and 09h50 - 14h50 at $\delta$ = +00$^\circ$. The
2dF instrument allows  spectra for  up to 400 QSO candidates to be obtained simultaneously. The input
catalogue is based on APM  UBR magnitudes for $7.5\times10^6$ stars to B=20.85 on 30 UKST fields. The
final QSO catalogue will be an order of magnitude bigger than previous complete  QSO surveys. The
prime  scientific aims of the 2dF QSO survey  are:

\begin{enumerate} \item To determine the QSO clustering power spectrum,  P(k),  in the range of
spatial scales, $0\la r\la 1000h^{-1}Mpc$.

\item To measure  $\Omega_\Lambda$  from  geometric distortions in clustering.

\item To trace the evolution of QSO clustering in the  range,  0$\la$z$\la$3,  to obtain new limits
on $\Omega_0$  and QSO bias.

\end{enumerate}

Other aims include determining the evolution of the QSO LF evolution to z=3, cross-correlating the
QSO's with 2dF Galaxy Survey  groups to measure $\Omega_0$ via gravitational lensing (cf. Croom \&
Shanks, 1999a) and   constraining  $\Omega_\Lambda$ by finding the sky density of close (6-20$''$)
lensed QSO pairs.

Previous QSO clustering results have generally been based on the  following complete QSO surveys -
the Durham/AAT survey of  Boyle et al (1988) comprising  392 B$<$21 QSO's, the CFHT survey of
Crampton et al (1989) with 215 B$<$20.5  QSO's, the ESO survey of Zitelli et al (1992) with 28 B$<$22
QSO's,  the LBQS survey of  Hewett et al (1995)  with 1053 B$<$18.8 QSO's and the ESO  survey of
LaFranca et al (1998) with 300 B$<$20.5 QSO's. These have produced results on QSO clustering at small 
(r$<10h^{-1}$ Mpc) scales which are consistent with an r$^{-1.8}$ power-law with amplitude
$r_0\approx6h^{-1}$Mpc which shows no evolution in comoving  coordinates (Shanks et al 1987, Andreani
\&  Cristiani, 1992, Shanks \& Boyle, 1994, Georgantopoulos \& Shanks, 1994,  Croom \& Shanks, 1996). 
At larger scales (10$<r<1000h^{-1}$Mpc), no  clustering has previously been detected  in $\xi_{qq}$.

\section{2dF QSO Redshift Survey Status and Current Results.}

We now have redshifts for $\approx$6000 QSO's where 5600 QSO's have been observed  using 2dF itself.
A further 400 QSO's selected from the input catalogue have been observed on different telescopes for
associated projects. Observations have included  bright 17$<B<$18.25 QSO's using UK Schmidt Telescope
FLAIR fibre coupler, radio-loud QSO's identified in the NRAO VLA Sky Survey (NVSS) and observed at 
Keck and finally 30 QSO's in close pairs ($<20''$) from the ANU 2.3-m  telescope. This makes the 2dF 
survey already the biggest, complete QSO survey by a factor of $\approx$6.

\begin{figure*}
\centering\mbox{\psfig{figure=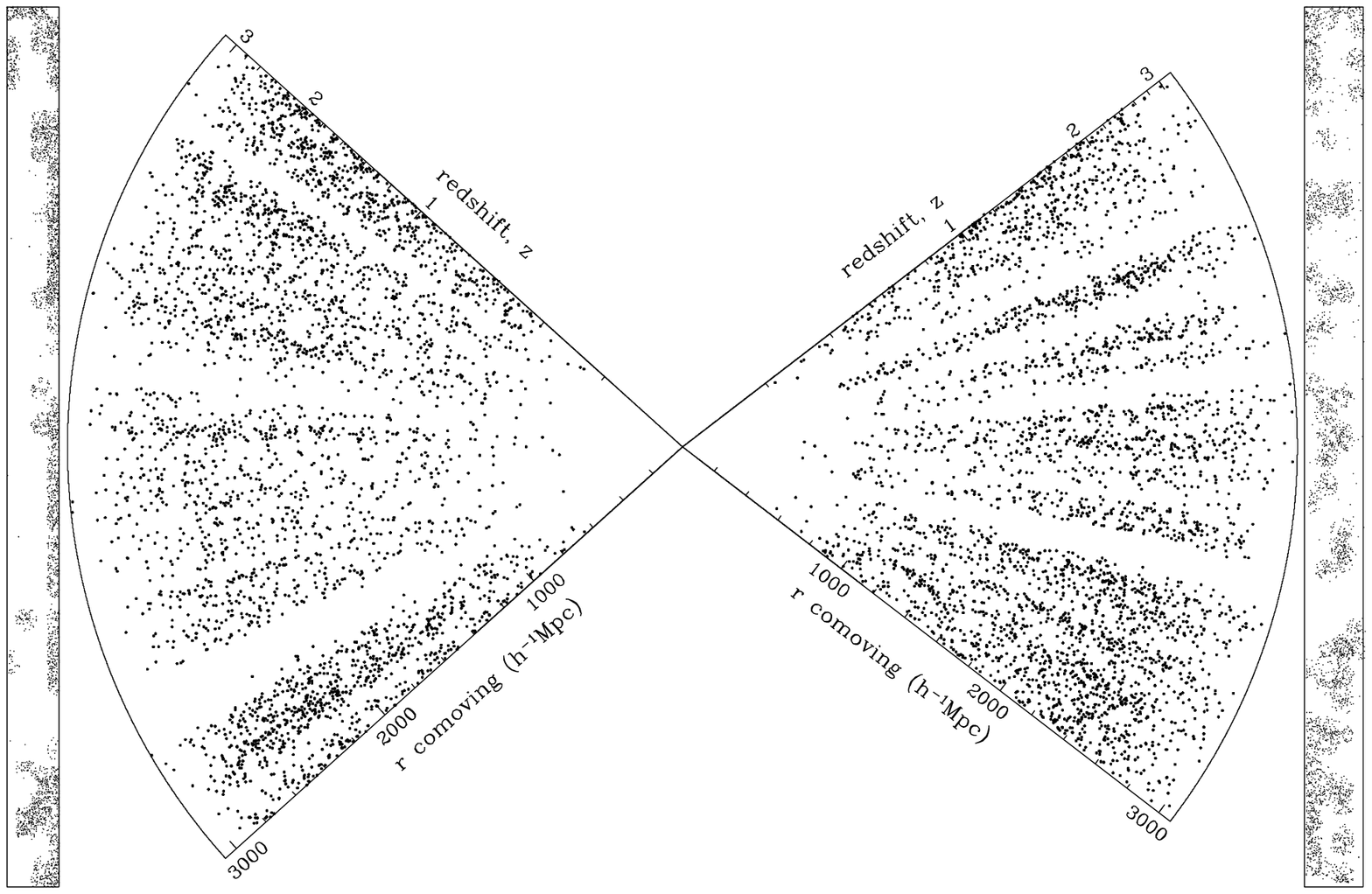,height=7cm,angle=-90}}
\caption[]{The distribution of $\approx$6000  QSO's with currently measured redshifts in the 2dF QSO
Redshift Survey. The right-hand wedge is the NGP area at $\delta=0^{\circ}$, while the left-hand
wedge is the SGP region at $\delta=-30^{\circ}$.   The rectangular strips show the survey's current
sky coverage.} 
\label{wedge} 
\end{figure*}

From these observations, we know that 53.5\% of candidates are QSO's which means there will be 26000
QSO's in  final survey. The QSO number count, n(B), has been found to be  in good agreement  with
previous surveys. The QSO redshift distribution, n(z), extends to z$\approx$3 because of our
multi-colour UBR selection. In Fig. \ref{wedge} we show the current Northern and Southern redshift
cone plots from the 2dF survey. The rectangles shows the sky distribution of  the fields that have
already been observed.

We have detected 8 close ($<20''$) QSO pairs. Only one is a candidate  gravitational lens; the
separation is $\approx16''$ (see Fig. 6 of Croom et al., 1999).

\begin{figure*} 
\centering\mbox{\psfig{figure=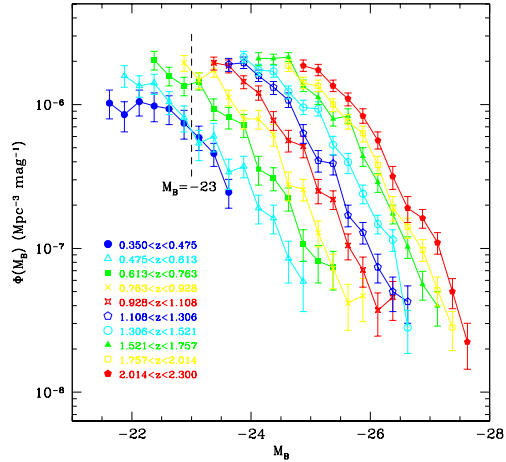,height=7cm}}
\caption[]{The 2dF + LBQS  QSO Luminosity Function based on $\approx$6000 QSO's for a q$_0$=0.5,
H$_0$=50kms$^{-1}$Mpc$^{-1}$ cosmology. Incompleteness at M$_B>-23$ is due to host galaxy
contamination.} 
\label{olf} 
\end{figure*}

The new 2dF results for  the QSO Luminosity Function  continue to  be consistent with Pure Luminosity
Evolution models throughout the range 0.35$<$z$<$2.3 (Boyle et al.,1988,1991,1999a). The luminosity
function based on the current $\approx$6000 QSOs of the 2dF survey  combined with $\approx$1000 LBQS
QSOs  are shown in Fig. \ref{olf}. The large sample size makes it possible to define the QSO
Luminosity Function in much smaller redshift bins than used previously and the accuracy of pure
luminosity evolution as a description of QSO evolution is clear. Indeed, this accuracy is so high
that it has prompted a new investigation of pure-luminosity evolution as a physical, as well as a
phenomenological, model for QSO evolution. (Done \& Shanks, 1999, in prep.)

The most interesting individual QSO that has been discovered from the 2dF QSO survey is UN
J1025-0040, a  unique, post-starburst radio QSO at z=0.634, identified in the 2dF-NVSS catalogue and
followed up spectroscopically  at the Keck (Brotherton et al., 1999). As well as broad emission
lines, the spectrum also shows strong Balmer absorption lines indicative of a post-starburst galaxy.
The starburst component of the spectrum at M$_B$=-24.7 dominates the AGN continuum spectrum by
$\approx$ 2mag. This 2dF-NVSS collaboration  has previously also uncovered a new class of radio-loud
BAL QSO's (Brotherton et al., 1998) and clearly has great potential for further exciting discoveries.

\begin{figure*} 
\centering\mbox{\psfig{figure=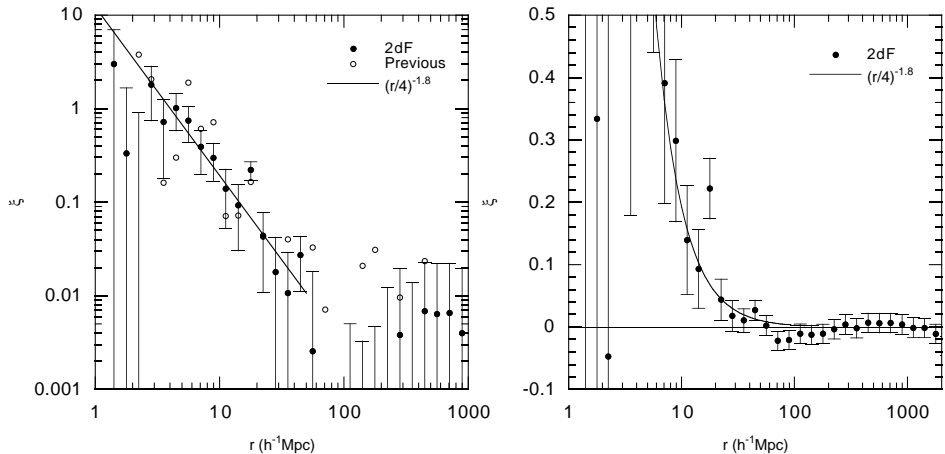,height=6cm}}
\caption[]{(a) The 2-point auto-correlation function for 4115 QSO's from the 2dF QSO redshift Survey
(closed circles), measured in comoving coordinates with q$_0$=0.5. The open circles are the combined
results from previous surveys (Croom \& Shanks, 1996). The line is a -1.8 power law with
$r_0=4h^{-1}$Mpc. (b) 2dF QSO correlation function as in (a). The line is a -1.8 power law with
$r_0=4h^{-1}$Mpc which fits out to $\approx$50h$^{-1}$Mpc.} 
\label{xir} 
\end{figure*}

Finally, we present a  preliminary  2dF QSO correlation function from our most complete subset of
4115 QSO's. We have taken into account the current incompleteness of the 2dF survey  as best we can;
however, this process is complicated by the fact that many observed areas still have overlapping 2dF
`tiles' as yet unobserved. We show the preliminary correlation function in Fig. \ref{xir}(a) as a
log-log plot and in Fig. \ref{xir}(b) as a log-linear plot. As can be seen the correlation function
is consistent with being a -1.8 power-law with $r_0\approx4h^{-1}$Mpc and in good agreement with
previous results (eg Croom and Shanks, 1996). The QSO correlation function  thus appears to be
remarkably similar to the correlation function for local, optically selected galaxies and continues
to show no evolution in comoving  coordinates. This behaviour is consistent with  evolution due to
gravitational clustering  either in a low $\Omega_0$ model or in a biased, $\Omega_0=1$ model. The
correlation function is consistently positive out to about 20h$^{-1}$Mpc at 3$\sigma$  and out to
about  50h$^{-1}$Mpc at 1$\sigma$, thus giving a hint of  power extending to larger scales. At
$r>50h^{-1}$Mpc the errors in $\xi$ are now as low as $\pm$0.016. There is also a suggestion  of
anti-correlation at r$\approx$100h$^{-1}$Mpc but still only at  1.5$\sigma$ significance. At all
scales in the range 100$<$r$<$1000h$^{-1}$Mpc the correlation function is within 1$\sigma$ of zero.

\section{Future Results.}

We have  used mock catalogues drawn from simulations using the Zeldovich  approximation to
demonstrate   the  accuracy of P(k)  estimates at large scales that will be obtainable from the full
survey   (Croom et al., in prep.). The results from this simulation show that our estimates of the
power spectrum can recover the input spectrum up to scales of several hundreds of Mpc
(k$\approx$0.01hMpc$^{-1}$) where the power spectrum is expected to turn over to its primordial form
(see Boyle et al., 1999b).

We are now using the Hubble Volume N-body simulation from the Virgo Consortium with of order 10$^9$
particles to produce mock QSO survey catalogues.  We have `light cone' output for a model with
$\Omega_{\Lambda}$=0.7, $\Omega_0$=0.3 and $\sigma_8$=1.0 to test the potential of  the geometric
test of Alcock \& Paczynski (1979) for $\Omega_{\Lambda}$ directly in a biased $\Lambda$CDM model
(Hoyle et al., 1999 in prep.). We shall be using these simulations to see how robust the test is
against different  models for the bias. Constraints on these bias models come from the QSO-galaxy
cross-corrrelation function  (Ellingson et al., 1991, Smith et al., 1995, Croom \&  Shanks, 1999b)
which suggest radio-quiet QSO's inhabit similar environments to average galaxies out to z=1.5. We are
now extending QSO environment tests out to z$\approx$2  using the INT Wide Field Camera at B$<$26 and 
the AAT Taurus Tunable Filter  (Croom \& Shanks, in prep.) and these results will further constrain
possible models of QSO bias at high redshift.

\section {Conclusions}

\begin{enumerate}

\item The 2dF QSO Survey has obtained redshifts for 6000/26000   B$<$20.85, z$<$3 QSOs which means it
is  already the biggest  complete QSO survey by a factor of $\approx$6.

\item Previous surveys detected QSO clustering at 4$\sigma$ level  for r$<$10h$^{-1}$Mpc where the
QSO clustering  appears stable when measured in comoving  coordinates, suggesting either a low
$\Omega_0$ model or a biased,  $\Omega_0$=1 model. Previous surveys detected no significant
clustering in $\xi_{qq}$ at larger scales.

\item The  2dF QSO survey confirms the accuracy of the PLE model of Boyle et al (1991) for the
evolution QSO Luminosity Function for 0$<$z$<$2.2.

\item The 2dF QSO survey has  already detected many individually interesting objects, including a
post-starburst QSO  and several close QSO  pairs.

\item A preliminary 2dF QSO correlation function based on 4115 QSOs shows results consistent with
previous  surveys and consistent with the correlation function of local, optically selected galaxies
but with a hint of extended power to $\approx$50h$^{-1}$Mpc.

\item Mock catalogues from the Hubble Volume suggest the 2dF QSO survey will be  able to determine
QSO power spectrum out to  $\ga$1000h$^{-1}$Mpc and so detect the expected turnover to the
primordial, n=1, slope.

\item The  potential of 2dF QSO  survey to constrain $\Omega_{\Lambda}$ from  both geometric
distortions and gravitational lensing  is being tested in  the Hubble Volume mock catalogues.

\end{enumerate}

\acknowledgments We thank Fiona Hoyle, Carlton Baugh and Adrian Jenkins for allowing us to use the
results from their analyses of the mock catalogues from the Hubble Volume simulation. The Hubble
Volume is provided courtesy of the Virgo Consortium.

\end{document}